\documentclass[9pt]{article}
\usepackage{listings}
\usepackage{amsthm}
\usepackage{amsmath}
\usepackage{amsfonts}

\usepackage{url}
\usepackage{graphicx}
\usepackage{color}

\definecolor{mygreen}{rgb}{0,0.6,0}
\definecolor{mygray}{rgb}{0.5,0.5,0.5}
\definecolor{mymauve}{rgb}{0.58,0,0.82}
\title{Median of heaps: linear-time selection by recursively constructing binary heaps}
\author{Oliver Serang\\
}

\begin{document}

\maketitle

\begin{abstract}
\noindent The first worst-case linear-time algorithm for selection was
discovered in 1973; however, linear-time binary heap construction was
first published in 1964. Here we describe another worst-case linear
selection algorithm, which is simply implemented and uses binary heap
construction as its principal engine. The algorithm is implemented in
place, and shown to perform similarly to in-place median of medians.
\end{abstract}

\flushbottom
\maketitle
\thispagestyle{empty}

\section{Introduction}
Selection is the problem of finding the $\tau^* = k^\text{th}$
smallest value in a collection. Online versions of this problem exist,
but or simplicity, here we focus on selecting from a list $x_1, x_2,
\ldots, x_n$. Selection can be trivially performed via sorting in
$O(n\log(n))$ to retrieve $\tau^*$. Selection can also be performed by
popping values from a heap or priority queue in $O(n +
k\log(n))$\cite{williams1964algorithm}; however, computing the median
(\emph{i.e.}, $k=\frac{n}{2}$) requires $O(n\log(n))$ steps.

Hoare's quickselect, published in 1961 algorithm was the first
algorithm with a runtime practically faster than sorting $\forall
k$.\cite{hoare1961quickselect} It was published concurrently with
Hoare's quicksort algorithm and operates in a similar
manner\cite{hoare1961quicksort}: Quickselect chooses a random element
$\tau=x_i$ and ``pivots'' by partitioning the array in $O(n)$ into
values $<\tau$ and values $\geq \tau$. In doing so, the rank of
$\tau$, $r(x,\tau) = |\{j : x_j<\tau\}|$ is counted. When
$k<r(x,\tau)$, recursive selection is continued on the smaller
values. Likewise, when $k>r(x,\tau)$, recursion is continued on the
larger values. And if $k=r(x,\tau)$, $\tau=\tau^*$.

While quickselect, like quicksort, can be seen to have an expected
runtime in $O(n\log(n))$ via a harmonic
series\cite{sedgewick1977analysis}, its worst-case runtime remains in
$\Omega(n^2)$. This can occur when $\tau\in\{min(x), max(x)\}$, or
more generally, $r(x,\tau)\in O(1) \wedge n-r(x,\tau)\in O(1)$ and
thus $\Theta(n)$ is spent to partition when only eliminating $O(1)$
elements.

In 1973, Blum \emph{et al.} discovered that selection could be
performed in worst-case $O(n)$ time\cite{blum1973time}. Blum et al.'s
``median of medians'' algorithm operates by grouping $x$ into
subgroups of length $5$. The median of each group of 5 is computed in
$O(1)$, for $O(n)$ steps over all groups of 5. The algorithm then
recurses to compute the median of these $\frac{n}{5}$ medians. This
median of medians value, $\tau$, is used for partitioning $x$ in the
same manner as quickselect; however, the median of medians is
proveably a middle element, where both $r(x,\tau)$ and $n-r(x,\tau)$
are provably large. This guarantees that when $\tau\neq\tau^*$, the
recursion on either the larger or smaller elements would consider only
a fraction of $n$. Specifically, $r(x,\tau),n-r(x,\tau) \geq
\frac{3n}{10}$, because each group of 5 where the median of the group
is $<\tau$ has 3 values $<\tau$, and there are $\approx\frac{n}{2}$
such groups. The resulting recurrence is $T(n) =
T\left(\frac{n}{5}\right) + T\left(\frac{7n}{10}\right) +
O(n)$. Together, the recursion to compute the median of medians and
the recursion to select the remainder consider at most
$\frac{9n}{10}$. This results in a root-heavy case of the master
theorem, wherein regularity guarantees $T(n)\in \Theta(n)$. Median of
medians is regarded as a groundbreaking algorithm for its approach.

In subsequent years, alternate algorithms have been discovered, which
select by traversing values in a heaps or by using approximate
priority queues. These algorithms are more sophisticated and
complicated to
implement\cite{frederickson1993optimal,chazelle2000soft}.

Since 1964, it was known that construction of a binary heap
(``heapifying'') can be performed by heapifying both the left and
right $\frac{n}{2}$ and then melding those heaps by reheaping in
$O(\log(n))$ operations\cite{williams1964algorithm}. The resulting
recurrence $T(n) = 2\cdot T\left(\frac{n}{2}\right) + \log(n)$ is a
leaf-heavy case of the master theorem, with $T(n)\in\Theta(n)$.

Here we introduce a fairly simple in-place and worst-case $O(n)$
selection algorithm, which follows directly from construction of a
binary heap itself. A partition element $\tau : r(x,\tau),n-r(x,\tau)
\geq c\cdot n$ for constant $c>0$ can be trivially discovered by
recursively heapifying one level of each heap tree and recursing
downward.

\section{Methods}
\subsection{Basic procedure}
The algorithm proceeds as follows:

First, heapify $x$ in an array, where $x_i<x_{2i+1},x_{2i+2}$. Levels
of the tree are contiguous in memory: level $i$ has $2^i$ values, $y_i
= x_{2^i}, x_{2^i+1}, \ldots, x_{2^{i+1}}-1$.

The heap is a complete tree. Let $d$ be the depth at which the heap is
perfect, $d = \lfloor \log_2(n + 1) - 1\rfloor$. Let $\ell$ be the
level of the tree where $\tau$ is found via recursive selection. Let
that recursive selection $\tau = select(y_\ell, k')$ partitions
$y_\ell$ via $\tau$ where $r(y_\ell,\tau)=k'-1$ elements in $y_\ell$
are $<\tau$ and $2^\ell-k'$ values are $\geq\tau$ (excluding the
element corresponding to $\tau$ itself).

\begin{figure}
  \centering
  \includegraphics[width=4.5in]{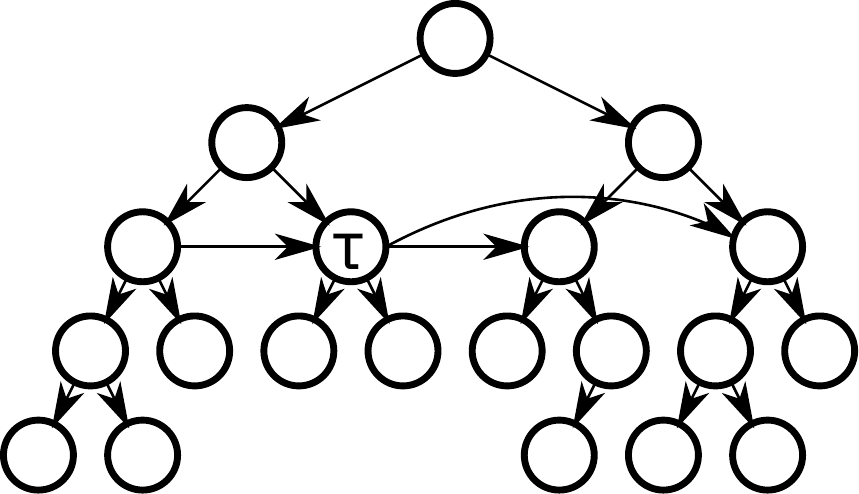}
  \caption{{\bf Illustration of stacked inequalities after selection
      on level $y_2$ of the heap tree.} A level toward the bottom of
    the tree, $y_2$, is recursively partitioned via selection to find
    $\tau$. The directed edge $a \rightarrow b$ depicts inequality
    $a<b$. Choosing a middle level in the tree results in stacked
    inequalities, represented here as chains of directed edges. This
    figure depicts $n=20$ values, $d=3$ deepest perfect layer, $t=5$
    remaining values in the final imperfect layer, and the $k'=1$
    value recursively selected from $y_2$.\label{fig:cartoon}}
\end{figure}

By choosing level $\ell<d$ and by choosing $k'$ to select somewhere
from the middle of $y_\ell$, inequalities can be stacked. This is
portrayed by chains in the digraph visualized by
Figure~\ref{fig:cartoon}: $y_\ell$ contains 1 value $<\tau$ and 2
values $\geq\tau$, but the 2 ancestors of all values $<\tau$ are
likewise $<\tau$. Similarly, the 7 descendents of values $\geq\tau$
are $\geq\tau$.

\subsection{Optimization of $k'$}
We will optimize both $k'$ and later the level $\ell$ from which it is
selected.

The $k'-1$ values in $y_\ell$ that are $<\tau$ imply at least $k'-1 +
\frac{k'-1}{2} + \frac{k'-1}{4} + \cdots + 2 + 1 = 2k'-3$ values
$<\tau$ (including the $k'+1$ nodes and their ancestors). The
$2^\ell-k'$ values $\geq\tau$ imply at least
$(2^\ell-k')\cdot(2^{d-\ell+1}-1)$ values $\geq\tau$.

We can balance the number of values $<\tau$ and $\geq\tau$ by allowing
$2k'-3 = (2^\ell-k')\cdot(2^{d-\ell+1}-1)$, which yields
\begin{eqnarray*}
  k' &=& \frac{2^{d+1} - 2^\ell + 3}{2^{d-\ell+1}+1}\\
  &\approx& \frac{2^{d-\ell+1} - 1}{2^{d-\ell+1} + 1} \cdot 2^\ell
\end{eqnarray*}

Because we've balanced the number of values provably $<\tau$ and
$\geq\tau$, at least $\approx 2k'$ values will be excluded. Hence, the
proportion of $x$ excluded will be $\approx\frac{2k'}{n}$.

The worst-case runtime is characterized by
\begin{eqnarray*}
  T(n) &=& T\left(2^\ell\right) + T\left( \left(1-\frac{2k'}{n}\right)\cdot n \right) + \Theta(n)\\
  &=& T\left( \frac{2^\ell}{n} \cdot n\right) + T\left( \left(1-\frac{2k'}{n}\right)\cdot n \right) + \Theta(n).
\end{eqnarray*}

Thus, regularity requires
\begin{eqnarray*}
  c &=& \frac{2^{\ell}}{n} + 1-\frac{2k'}{n}\\
  &=& 1 + \frac{2^{\ell} - 2k'}{n}\\
  &=& 1 + 2^\ell\cdot\frac{1 - 2\cdot\frac{2^{d-\ell+1} - 1}{2^{d-\ell+1} + 1}}{n} < 1. 
\end{eqnarray*}

\subsubsection{Case 1: perfect heap}
When the heap is perfect, $n=2^{d+1}-1$; therefore,
\[
  c_{\text{perfect}} \approx 1 + 2^{\ell-d-1}\cdot \left(1 - 2\cdot\frac{2^{d-\ell+1} - 1}{2^{d-\ell+1} + 1}\right).
\]

\subsubsection{Case 2: imperfect heap}
When the heap is imperfect, $n \in \left(2^{d+1}-1, 2\cdot(2^{d+1}-1)
\right) \approx \left(2^{d+1}, 2^{d+2}\right).$ Thus,
\[
  c_{\text{imperfect}} \approx 1 + \lambda\cdot 2^{\ell-d-1}\cdot \left(1 - 2\cdot\frac{2^{d-\ell+1} - 1}{2^{d-\ell+1} + 1}\right),
\]
where $\lambda\in \left(\frac{1}{2},1\right)$.

Assuming regularity in case 1, the worst-case $c_{\text{imperfect}}$
will occur at $\lambda=\frac{1}{2}$, because $c_{\text{perfect}}$
decreases from 1, where minimizing $\lambda$ will minimize this
decrease from 1. First, this proves that regularity in case 1 implies
regularity in case 2, and would thus imply $T(n)\in
\Theta(n)$. Second, it yields the worst-case $c$ for imperfect heaps:
\[
  c_{\text{imperfect}} \approx 1 + 2^{\ell-d-2}\cdot \left(1 - 2\cdot\frac{2^{d-\ell+1} - 1}{2^{d-\ell+1} + 1}\right).
\]

\subsubsection{Case 3: revising the algorithm to tighten the worst-case bound on imperfect trees}
The runtime bound above for imperfect heaps is conservative.

Let $t=n - 2^{d+1}-1 \geq 0$ be the number of nodes in the final,
imperfect level of the heap if the tree. When the tree is perfect,
$t=0$. At worst, $k'\cdot (2^{d-\ell+1}-1)$ of the $t$ values in the
imperfect level will have non-stacking inequalities. \emph{I.e.}, at
worst, these values' ancestors in $y_\ell$ are $<\tau$, but the values
are $\geq$ their ancestors, leaving non-stacking inequalities. The
remaining at least $t - k'\cdot (2^{d-\ell+1}-1)$ values that must be
$\geq\tau$. Thus, we could achieve tighter bounds by solving
\begin{eqnarray*}
2k' &\approx& (2^\ell-k')\cdot(2^{d-\ell+1}-1) + t - k'\cdot(2^{d-\ell+1}-1)\\
\leftrightarrow k' &\approx& \frac{2^{d+1}+t-2^\ell}{2^{d-\ell+2}}\\
&\approx& \frac{n-2^\ell}{2^{d-\ell+2}}.
\end{eqnarray*}

Once again, let $\lambda \approx \frac{2^{d+1}}{n} \in
\left[\frac{1}{2},\lambda_{\max}\right)$, where $\lambda_{\max}<1$ is
  determined by $t - k'\cdot(2^{d-\ell+1}-1) = 0$.

\begin{eqnarray*}
  c &=& 1 + \frac{2^{\ell} - 2k'}{n}\\
  &=& 1 + \frac{2^\ell}{n} - 2\cdot\frac{1 - \frac{2^\ell}{n}}{2^{d-\ell+2}}.
\end{eqnarray*}
$\frac{2^\ell}{n} = \lambda\cdot 2^{\ell-d-1}$; therefore, denoting
$a=2^{\ell-d-1}\in(0,1)$, we have
\begin{eqnarray*}
  c &=& 1 + \lambda\cdot a - 2\cdot\frac{1 - \lambda\cdot a}{\frac{2}{a}}\\
  &=& 1 + \lambda\cdot a - a + \lambda\cdot a^2\\
  &=& 1 + a\cdot (\lambda\cdot(1+a) - 1).
\end{eqnarray*}
$\frac{\partial c}{\partial \lambda}\neq 0$, so the worst-case
$\lambda$ occurs at a boundary $\lambda=\lambda_{\max}$. Thus we see
that the worst-case scenario is achieved by $t -
k'\cdot(2^{d-\ell+1}-1) = 0$.

When this occurs, we choose $k'$ as in cases 1 \& 2. In bounding $c$,
what changes is
\begin{eqnarray*}
  \frac{2^\ell}{n} &=& \frac{2^\ell}{2^{d+1}-1+t}\\
  &\approx& \frac{2^\ell}{2^{d+1}+t}\\
  &=& \frac{2^\ell}{2^{d+1}+k'\cdot(2^{d-\ell+1}-1)}\\
  \rightarrow c &=& 1 + 2^\ell\cdot\frac{1 - 2\cdot\frac{2^{d-\ell+1} - 1}{2^{d-\ell+1} + 1}}{2^{d+1}+k'\cdot(2^{d-\ell+1}-1)}\\
  &=& 1 + 2^\ell\cdot\frac{1 - 2\cdot\frac{2^{d-\ell+1} - 1}{2^{d-\ell+1} + 1}}{2^{d+1}+k'\cdot(2^{d-\ell+1}-1)}\\
  &=& 1 + \frac{1 - 2\cdot\frac{2^{d-\ell+1} - 1}{2^{d-\ell+1} + 1}}{2^{d-\ell+1}+\frac{k'}{2^\ell}\cdot(2^{d-\ell+1}-1)}\\
  &=& 1 + \frac{1 - 2\cdot\frac{2^{d-\ell+1} - 1}{2^{d-\ell+1} + 1}}{2^{d-\ell+1}+\frac{2^{d-\ell+1} - 1}{2^{d-\ell+1} + 1}\cdot(2^{d-\ell+1}-1)}.
\end{eqnarray*}

Listing~3 implements this revised algorithm.

\subsubsection{Optimizing $\ell$ to minimize worst-case $c$}
Table~\ref{tab:ell-c} shows the relationship between $\ell$ and $c$
for perfect and imperfect heaps. The best asymptotic worst-case
runtime (corroborated analytically, not shown) is given by
$\ell=d-2$. For the basic median of heaps, $k'=\frac{7}{9}\cdot
2^\ell$, achieving $c<1$. This generalizes the runtime bound to prove
$T(n)\in\Theta(n)$ for the basic median of heaps algorithm for both
perfect and imperfect heaps.

The revised algorithm yields a tighter bound on $c$ for imperfect
heaps. Note that although both the basic and revised algorithms behave
identically on the first recursion when $n=2^i-1$ (\emph{i.e.}, a
perfect heap), the recursion after partitioning is no longer
guaranteed to be perfect. Thus, the worst-case constant of geometric
decay will be determined by $c_{\text{imperfect}}$. 

\begin{table}
  \centering
  \footnotesize
\begin{tabular}{r|ccc}
  $d-\ell$ & $c_{\text{perfect}}$ & $c_{\text{imperfect}}$ (basic alg.) & $c_{\text{imperfect}}$ (revised alg.)\\
\hline
  0 & 1.1667 & 1.0833 & 1.1429\\
  1 & 0.9500 & 0.9750 & 0.9655\\
  2 & 0.9306 & 0.9653 & 0.9587\\
  3 & 0.9522 & 0.9761 & 0.9738\\
  4 & 0.9725 & 0.9863 & 0.9856\\
  5 & 0.9853 & 0.9927 & 0.9925
\end{tabular}
\caption{{\bf Relationship between $\ell$ and $c$.} For $k'$ chosen as
  described, the worst-case constant of geometric decay for the median
  of heaps algorithm is shown for perfect heaps. Also shown are the
  worst-case gemoetric decay constants for the basic and revised
  median of heaps algorithms on imperfect heaps. In all cases, the
  greatest decay in the worst case is found $\ell=d-2$. Values are
  rounded to ${10}^{-4}$.\label{tab:ell-c}}
\end{table}

\subsection{Implementation}
Quickselect was implemented via random pivot and {\tt std::partition}.

Median of medians was implemented in place by striding by powers of 5
as recursions become deeper. Strided arrays are accessed via a class
and iterators, allowing use of {\tt std::sort} on each group of 5 as
well as selecting with {\tt std::sort} when $n<16$. When $n \not\equiv
0 \text{(mod 5)}$, the remaining 4 or fewer values were excluded when
computing the median of medians.

Listing~1 and Listing~2 demonstrate implementations in {\tt python}
and {\tt C++}, respectively. Neither implementation considers $t$ when
choosing $k$, deferring to the weaker bound for imperfect heaps. For
expected rather than worst-case performance, the implementations use
$\ell=d-1$ and thus $k'=\frac{3}{5}\cdot 2^\ell$.

By changing to another selection algorithm when $n$ is a sufficiently
large constant base case, we are assured the existence of level
$\ell=d-1$ in the heap. Here we stop recursion when $n<16$, which at
most corresponds to a perfect heap of $8+4+2+1 = 15$ values.

Listing~3 shows the more complicated revised median of heaps
algorithm, achieving a tighter worst-case bound on $c$.

\clearpage
\begin{lstlisting}[language=Python, firstnumber=1, caption={{\bf Python implementation of basic median of heaps algorithm.} Using $\ell=d-1$ and thus $k'=\frac{3}{5}\cdot 2^\ell$.\label{alg:median-of-heaps-python}}]
import numpy as np
from heapq import heapify
import math

def pivot(x, tau):
  x = np.array(x)
  return list(x[x<tau]), list(x[x>=tau])

def sel_sort(x, k):
  return sorted(x)[k]

def sel_median_of_heaps(x, k):
  n = len(x)
  if n < 16:
    return sorted(x)[k]

  x = list(x)
  heapify(x)
  d = int(math.log2(n+1)-1)

  # second-to-last perfect level in the tree:
  y = x[2**(d-1)-1 : 2**(d-0)-1]

  tau = sel_median_of_heaps(y, 3*len(y)//5)

  x_small, x_big = pivot(x, tau)
  n_small = len(x_small)
  if k < n_small:
    return sel_median_of_heaps(x_small, k)
  return sel_median_of_heaps(x_big, k-n_small)
\end{lstlisting}
\clearpage

\begin{lstlisting}[language=C++, caption={{\bf In-place {\tt C++} implementation of median of basic heaps algorithm.} Using $\ell=d-1$ and thus $k'=\frac{3}{5}\cdot 2^\ell$.\label{alg:median-of-heaps-c}}]
#include <algorithm>
#include <functional>
#include <math.h>


template <typename T>
T sel_sort(T*x, long n, long k) {
  std::sort(x, x+n);
  return x[k];
}

template <typename T>
T sel_median_of_heaps(T*x, long n, long k) {
  if (n<16)
    return sel_sort(x, n, k);

  /* builds max heap, so use > as < to build a min heap: */
  std::make_heap(x, x+n, std::greater<T>());

  /* final perfect level in the tree: */
  long d = log2(n+1) - 1;
  /* second-to-last perfect level in the tree: */
  long i_start = (1L<<(d-1)) - 1;
  T*y = x + i_start;
  long len_y = 1L<<(d-1);

  long tau = sel_median_of_heaps(y, len_y, 3*len_y/5);

  /* partition into the values < tau and the values >= tau: */
  T*x_prime = std::partition(x, x+n, [tau](const auto & val) {
      return val<tau; });

  long n_small = x_prime - x;
  if (k < n_small)
    return sel_median_of_heaps(x, n_small, k);
  return sel_median_of_heaps(x_prime, n-n_small, k-n_small);
}

\end{lstlisting}
\clearpage

\begin{lstlisting}[language=C++, caption={{\bf In-place {\tt C++} implementation of median of heaps revised to tighten $c$ bound.} Using $\ell=d-1$. This algorithm is shown for demonstration and is not benchmarked, because performance in practice is too similar to the simpler median of heaps variant.\label{alg:median-of-heaps-c-2}}]
template <typename T>
T sel_median_of_heaps_tight(T*x, long n, long k) {
  // increased from 16 because approx has been applied sufficiently
  // that bounds need a slightly larger base case.
  if (n<32)
    return sel_sort(x, n, k);

  // makes a max heap, so use > as the < operator to build a min heap:
  std::make_heap(x, x+n, std::greater<T>());

  // final perfect level in the tree:
  long d = log2(n+1) - 1L;

  long ell = d-1L;
  
  long len_y = 1L<<ell;
  long i_start = len_y - 1L;
  T*y = x + i_start;

  // number of nodes in the final, imperfect layer:
  long t = n - ((1L<<(d+1)) - 1L);
  // solve 2k = (2**ell - k)*(2**(d-ell+1) - 1) + t - k*(2**(d-ell+1) - 1)
  long k_prime = (n - len_y) / (1L<<(d-ell+2L));

  if ( t - k_prime*( (1L<<(d-ell+1L)) - 1L ) < 0) {
    // final, imperfect layer has few values. revert to basic version
    // and solve 2k = (2**ell - k)*(2**(d-ell+1) - 1)
    long fold_vs_last_full = 1L<<(d-ell+1);
    k_prime = (fold_vs_last_full - 1) * len_y / (fold_vs_last_full + 1);
  }

  long tau = sel_median_of_heaps_tight(y, len_y, k_prime);

  // partition into the values < tau and the values >= tau:
  T*x_prime = std::partition(x, x+n, [tau](const auto & val) {
    return val<tau;});

  long n_small = x_prime - x;
  if (k < n_small)
    return sel_median_of_heaps_tight(x, n_small, k);
  return sel_median_of_heaps_tight(x_prime, n-n_small, k-n_small);
}
\end{lstlisting}

\section{Results}
\begin{figure}
  \centering
  \includegraphics[width=4.6in]{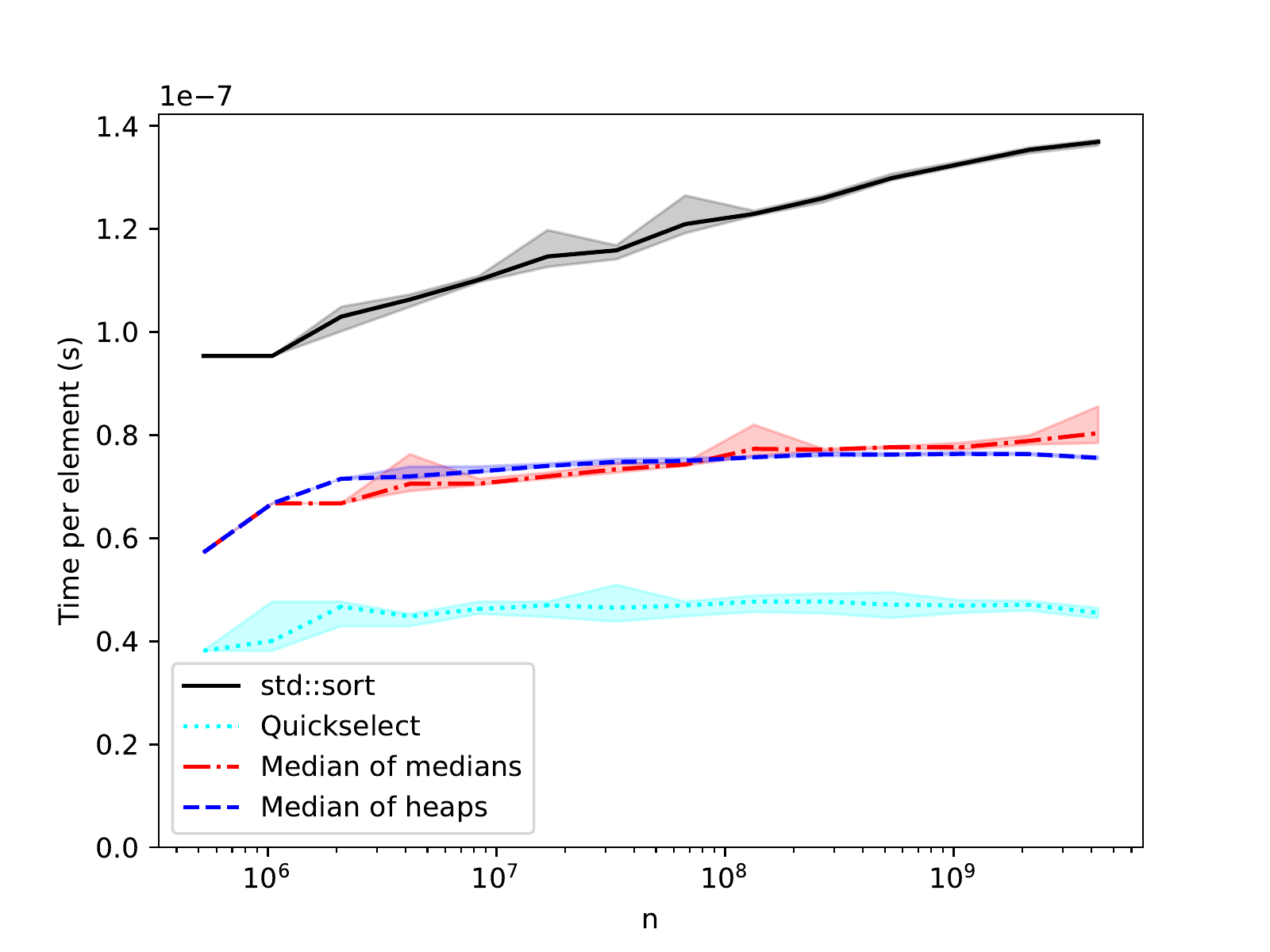}
  \caption{{\bf Per-element runtime of selection algorithms in {\tt
        C++}.} An $n$-element array is allocated, filled with random
    values, and the median is selected via $k=\frac{n}{2}$. Real
    elapsed time, including initialization of the array $x$, is
    recorded and aggregated over 5 replicate trials. Per-element
    runtimes compute the executable runtime divided by $n$. At each
    $n$, average runtime for an algorithm over the 5 replicates is
    plotted. The min and max over the 5 replicates are shaded. Note
    that for median of heaps, the runtime of the basic (not revised)
    algorithm is shown.\label{fig:runtimes}}
\end{figure}

{\tt C++} runtimes were computed with $n\in \{2^{19}, 2^{20}, \ldots,
2^{32}\}$. Compilation was performed using {\tt gcc 11.3.0} and with
flags {\tt -std=c++20 -O3 -march=native} an Intel {\tt i9-10885H} CPU
at 2.40GHz and with 64GiB RAM. At each $n$, selection of the median
($k=\frac{n}{2}$) is performed via 4 algorithms: {\tt std::sort},
quickselect, median of medians, and median of heaps (the simple
variant, not variant with the tight bound on $c$). 5 replicate trials
were timed for each $n$ and each algorithm. $x$ was filled with 64-bit
random integers via {\tt(rand() << 20) \^{} rand()}. Random values
were generated $\in \{0, 1, 2, \ldots, 32 n - 1\}$. On a given
replicate, each algorithm was timed using the same random seeds via
{\tt srand} to ensure the algorithms are compared on the same
inputs. Mean, min, and max runtimes are depicted in
Figure~\ref{fig:runtimes}. Mean runtimes are reported in
Table~\ref{tab:runtimes}.

\begin{table}
  \centering
  \scriptsize
\begin{tabular}{rrrrr}
         n &  std::sort &  Quickselect &  Median of medians &  Median of heaps (basic) \\
\hline
    524288 &      0.050 &        0.020 &              0.030 &            0.030 \\
   1048576 &      0.100 &        0.042 &              0.070 &            0.070 \\
   2097152 &      0.216 &        0.098 &              0.140 &            0.150 \\
   4194304 &      0.446 &        0.188 &              0.296 &            0.302 \\
   8388608 &      0.924 &        0.388 &              0.592 &            0.612 \\
  16777216 &      1.924 &        0.788 &              1.208 &            1.242 \\
  33554432 &      3.888 &        1.560 &              2.462 &            2.510 \\
  67108864 &      8.118 &        3.150 &              4.986 &            5.034 \\
 134217728 &     16.502 &        6.400 &             10.380 &           10.162 \\
 268435456 &     33.820 &       12.804 &             20.718 &           20.468 \\
 536870912 &     69.740 &       25.284 &             41.692 &           40.928 \\
1073741824 &    142.420 &       50.380 &             83.382 &           81.998 \\
2147483648 &    290.772 &      101.070 &            169.420 &          163.908 \\
4294967296 &    588.102 &      195.430 &            345.214 &          324.660 \\
\hline
\end{tabular}
\caption{{\bf Runtimes of selection algorithms in {\tt C++}.} Mean
  runtimes are rounded to $10^{-3}$ seconds, computed as described in
  Figure~\ref{fig:runtimes}. Note that for median of heaps, the
  runtime of the basic (not revised) algorithm is
  shown.\label{tab:runtimes}}
\end{table}

\section{Discussion}

Figure~\ref{fig:runtimes} shows gap growing linearly between {\tt
  std::sort} and the three selection algorithms benchmarked. This gap
without bound for $n\gg 1$ indicates different in-practice $O(\cdot)$
runtime.

The proposed median of heaps algorithm is roughly comparable to median
of medians. Both algorithms' costs per element plateau, and are
bounded above with the exception of slightly decreasing cache
performance as $n\gg 1$. The more complex revised median of heaps
variant performs nearly indistinguishably in practice (results not
shown). It is doubtful whether the performance gain in practice
justifies the additional complexity.

Both the basic and revised median of heaps algorithms here are
conservative in performing comparisons: during partitioning, the
algorithm partitions values whose ranks relative to $\tau$ are already
known (\emph{i.e.}, those left and above or right and below $\tau$ in
Figure~\ref{fig:cartoon}). Proceeding out-of-place and using a buffer
into which elements are copied may improve performance.

Both the median of medians and median of heaps algorithms are slower
here than quickselect. This is not unexpected, as quickselect is
lightweight and with low constant in expected
performance.\cite{sedgewick1977analysis}

On practical performance, access is sequential during partitioning and
thus cache performance is essentially optimal; however, reheaping
during heapify accesses children $2i+1,2i+2$, which can be far from
$i$ when $n\gg 1$, particularly during reheap propagation, when
accessing parents and further ancestors does not access levels of the
tree sequentially.

Similar cache penalties are incurred in the strided median of medians
implementation. This could likewise be buffered, copying the medians
to the buffer to recurse in-place. Buffering median of medians would
have cache benefits, but would result in greater memory footprint and
more copying of data. 

Interestingly, the recursive calls of median of heaps unroll to many
heapify operations on increasingly fine collections of values. The
worst-case linear runtime implies good $\tau$ candidates can always be
found in certain levels of the heap. These middle levels are enriched
for these middle $\tau$ values. Other options for finding these middle
$\tau$ values could be considered, thereby avoiding the recursion to
compute $\tau$.

\section{Conclusion}
The proposed median of heaps algorithm is a simple alternative for
solving selection in worst-case linear time and runtime in practice
similar to median of medians. The proposed algorithm uses a strategy
similar to median of medians. It is unclear if this approach would be
as accessible without first seeing Blum \emph{et al.}'s 1973 median of
medians algorithm; however, the proposed algorithm is concisely
implemented by using heapify and partition operations. It uses only
approaches known in the era when the complexity of selection was yet
unknown.

The algorithm is mainly a curiosity, but has been hybridized with
quickselect for good practical performance and guaranteed $O(n)$
worst-case (not shown here).


\end{document}